\documentclass[12pt]{article}
\usepackage{times}
\usepackage{amssymb,latexsym}
\usepackage{amsmath}
\usepackage{graphicx}
\usepackage{graphics}
\usepackage{extarrows}
\usepackage{float}
\usepackage{cite}
\usepackage{epsfig}
\usepackage{subfig}
\usepackage{mathrsfs}
\usepackage{fancyhdr}
\usepackage{tensor}
\usepackage{multicol}
\usepackage{setspace}
\usepackage[usenames]{color}
\usepackage[left=2.5cm,top=3cm,bottom=2.5cm,right=2.5cm,columnsep=.6cm]{geometry}

\newcommand{\al}[1]{\begin{align}#1\end{align}}
\newcommand{\bs}{\begin{split}}
\newcommand{\es}{\end{split}}
\newcommand{\eqr}{\eqref}
\newcommand{\mc}{\mathcal}
\newcommand{\pa}[1]{\left(#1\right)}
\newcommand{\pb}[1]{\left[#1\right]}

\numberwithin{equation}{section}

\usepackage[colorlinks=true,linkcolor=blue,citecolor=blue,urlcolor=blue]{hyperref}
\usepackage[nameinlink]{cleveref}

\begin{document}

%-------------------------------------------------------------------------------------------------------------------------------------------

\title{\vspace{-2cm} \textbf{Topologically Massive Yang-Mills Theory \\ and Link Invariants}}
\author{T. YILDIRIM\footnote{\href{mailto:tuna-yildirim@uiowa.edu}{tuna-yildirim@uiowa.edu}}
\vspace{.3cm}\\ 
\textit{Department of Physics and Astronomy,} \\
\textit{The University of Iowa, Iowa City, IA 52242}
\date{\normalsize\today}}

\maketitle
\vspace{-1.3cm}

%-------------------------------------------------------------------------------------------------------------------------------------------

\begin{center}
\noindent\line(1,0){415}
\end{center}

\begin{abstract}
\noindent
Topologically massive Yang-Mills theory is studied in the framework of geometric
quantization. Since this theory has a mass gap proportional to the topological mass $m$, Yang-Mills contribution decays exponentially at very large distances compared to $1/m$, leaving a pure Chern-Simons theory with level number $k$. In this paper, the \emph{near} Chern-Simons limit is studied where the distance is large enough to give an almost topological theory, with a small contribution from the Yang-Mills term. It is shown that this almost topological theory consists of two copies of Chern-Simons with level number $k/2$, very similar to the Chern-Simons splitting of topologically massive AdS gravity. Also, gauge invariance of these half-Chern-Simons theories is discussed.  As $m$ approaches to infinity, the split parts add up to give the original Chern-Simons term with level $k$. Reduction of the phase space is discussed in this limit. Finally, a relation between the observables of topologically massive Yang-Mills theory and Chern-Simons theory is shown. One of the two split Chern-Simons pieces is shown to be associated with Wilson loops while the other with 't Hooft loops. This allows one to use skein relations to calculate topologically massive Yang-Mills theory observables in the near Chern-Simons limit.
\end{abstract}

\begin{center}
\noindent\line(1,0){415}
\end{center}

%-------------------------------------------------------------------------------------------------------------------------------------------
%-------------------------------------------------------------------------------------------------------------------------------------------

\section{Introduction}\label{sec:intro}

Chern-Simons(CS) theory has been extensively studied and is a very important part of mathematical physics, mostly because of its connection with the link invariants of knot theory. This was first demonstrated by Witten \cite{Witten:1988hf} using 2D conformal field theories related to CS theory. Witten showed that Wilson loop expectation values(WLEV) of CS theory are given by link invariant polynomials which can be recursively calculated from skein relations. Later, Cotta-Ramusino et al. \cite{CottaRamusino:1989rf} derived skein relations for CS theory using only 3D field theory techniques. No matter what method is used, one crucial requirement for relating WLEV to link invariants is that the action must be metric free. Thus, it is not possible to do this with a Yang-Mills(YM) action. The main goal of this paper is to find out how CS link invariants are modified with the presence of a YM term at large but finite distances, where the metric contribution is very small, hence the theory is almost topological.

It is well known that adding a CS term to YM action in 2+1 dimensions gives mass to gauge bosons\cite{schonfeld1981mass}. A considerable amount of work on topologically massive Yang-Mills(TMYM) theory was done by Deser, Jackiw and Templeton\cite{Deser1982372,Deser2,Deser1984371} in the early 80s and some of the later work can be found in refs. \citen{Gonzales1986104, horvathy, Evens, martinez1989constrained, Giavarini1992222, Giavarini1993qd, Asorey1993477, Grignani1997360, Karabali:1999ef, dunne2000magnetic, Canfora2013}. Despite TMYM theory is more complicated than CS theory, there are surprising similarities between these theories, partly because conjugate momenta of both theories are given by gauge fields. An interesting example that shows the similarity of these two theories is the classical equivalence, first observed by Lemes et al\cite{Lemes:1997vx,Lemes:1998md}. This equivalence shows that classically it is possible to write the TMYM action as a pure CS action via a non-linear redefinition of the gauge fields. The Jacobian of this redefinition is 1 up to first order in the inverse topological mass expansion. Thus, at least for a large topological mass, one would expect that it might be possible to extend this equivalence to quantum level, but we will show that phase space geometry does not allow this. Instead, we will obtain a more complicated equivalence between the observables of both theories at large but finite distances.

Since pure Yang-Mills theory in 2+1 dimensions has a mass gap\cite{Karabali:1996iu}, the theory is trivial at very large distances. However, in TMYM theory that is not the case. The mass gap of this theory is proportional to the topological mass $m$\cite{Karabali:1999ef}. Studying large values of topological mass is equivalent to scaling up the metric or looking at large distances. In this paper, we study the theory at large but finite distances by neglecting the second and higher order terms in $1/m$, while keeping the first. This leads to an almost topological theory with a small contribution from YM, which allows us to write TMYM observables in terms of WLEVs of CS theory. This means that, not only in the pure CS limit but also in the \emph{near} CS limit, one only needs skein relations to calculate TMYM observables.

To show how the observables of CS and TMYM theories are related, we will start by reviewing Bos and Nair's work\cite{Bos:1989kn} on geometric quantization of CS theory in \autoref{sec:cs}. In \autoref{sec:wilcs}, we will briefly discuss the Wilson loops of CS theory. In \autoref{sec:tmym}, we will use similar methods on TMYM theory to find the wave functional and the gauge invariant integral measure. \Cref{sec:TMG} is a short discussion on topologically massive AdS gravity, which exhibits an interesting behavior analogous to our result. Then in \autoref{sec:wilson}, we discuss the link invariants of TMYM theory.

%-------------------------------------------------------------------------------------------------------------------------------------------
%-------------------------------------------------------------------------------------------------------------------------------------------

\section{Chern-Simons Theory}\label{sec:cs}

In this section, we would like to review the geometric quantization of non-Abelian CS theory, following Bos and Nair's work \cite{Bos:1989kn,Nair:2005iw}\footnote{Another comprehensive discussion on this subject can be found in ref. \citen{axelrod}}. Later, we will do a similar analysis on TMYM theory. 

The CS action is given by
\al{
S_{CS}=-\frac{k}{4\pi} \int \limits_{\Sigma\times[t_i,t_f]} {d^3x}\ \epsilon^{\mu\nu\alpha}\ Tr \pa{A_\mu \partial_\nu A_\alpha + \frac{2}{3}A_\mu A_\nu A_\alpha}
}
where $\Sigma$ is an orientable two dimensional surface. This action is classically not gauge invariant, but in the quantum theory it can be made gauge invariant by restricting k to be an integer.
The equations of motion for this theory are
\al{
F_{\mu\nu}=\partial_\mu A_\nu-\partial_\nu A_\mu+[A_\mu,A_\nu] =0.
}
Here, $A_{\mu}=-iA^a_{\mu}t^a$ where $t^a$ are matrix representatives of the Lie algebra $[t^a,t^b]=if^{abc}t^c$ and in the fundamental representation they are normalized as $Tr(t^at^b)=\frac{1}{2}\delta^{ab}$.
In the temporal gauge $A_0=0$ with $A_z=\frac{1}{2}(A_1+ iA_2)$ and $A_{\bar{z}}=\frac{1}{2}(A_1- iA_2)$, the CS action becomes
\al{
S_{CS}=-\frac{ik}{2\pi}\int{dt d\mu_\Sigma }\ Tr(A_{\bar z}\partial_0 A_z - A_{z}\partial_0 A_{\bar z} ).
}
The equations of motion in this gauge makes $A_z$ and $A_{\bar{z}}$ time independent along with constraining $F_{z\bar{z}}=0$. A very important feature is that the conjugate momenta are given by the gauge fields,
\al{
\Pi^{a z}=\frac{ik}{4\pi} A^a_{\bar{z}}~\ \text{and}~\ \Pi^{a \bar{z}}=-\frac{ik}{4\pi}  A^a_z.
}
Later, we will see a similar behavior in TMYM theory which is crucial for our work.

The symplectic two-form of CS theory is given by
\al{
\label{eq:omegacs}
\Omega=\frac{ik}{2\pi}\int \limits_{\Sigma}\delta A^a_{\bar{z}}\delta A^a_z.
}
In simply connected spaces it is possible to parametrize the gauge fields as 
\al{
\label{eq:par1}
A_{\bar z}=-\partial_{\bar z}UU^{-1}~\text{and}~A_z=U^{\dagger-1}\partial_z U^{\dagger}.
}
Here, U is a complex SL(N,$\mathbb{C}$) matrix which transforms like $U^g=gU$  where $g\in\mc{G}$ and $\mc{G}$ is the gauge group. We will continue with taking $\mc{G}=SU(N)$. $U$ is given by
\al{
\label{eq:U}
U(x,0,C)=\mc{P}exp\pa{-\underset{C}{\ \ \int_0^x}(A_{\bar{z}}d\bar{z}+\mc{A}_zdz)},
}
where $\mc{A}_z$ satisfies $\partial_z A_{\bar{z}}-\partial_{\bar{z}}\mc{A}_z+[\mc{A}_z,A_{\bar{z}}]=0$ and this flatness condition makes $U$ invariant under small deformations of the path $C$ on $\Sigma$.  From \eqref{eq:U}, it follows that
\al{
\label{eq:scriptA}
\mc{A}_z=-\partial_z U U^{-1} ~\text{and}~ \mc{A}_{\bar{z}}= U^{\dagger-1}\partial_{\bar{z}} U^{\dagger}.
}

%-------------------------------------------------------------------------------------------------------------------------------------------

\subsection{The Wave-Functional}

We choose the K\"ahler polarization which makes the quantum wave-functional $\psi$ only $A_{\bar{z}}$ dependent. The pre-quantum and quantum wave-functionals can then be related by $\Phi[A_z,A_{\bar{z}}]=e^{-\frac{1}{2}K}\psi[A_{\bar{z}}]$, where $K=\frac{k}{2\pi}\int_{\Sigma} A^a_{\bar{z}}A^a_z$ is the K\"ahler potential. Since the phase space is K\"ahler, the pre-quantum inner product can be retained at the quantum level\cite{Nair:2005iw}\footnote{A detailed discussion on geometric quantization can be found in ref. \citen{hall2013}} as,
\al{
\langle 1|2 \rangle =\int d\mu(\mc{M})\Phi_1^*\Phi_2 ~\rightarrow~ \int d\mu(\mc{M})e^{-K}\psi_1^*\psi_2
}
where $d\mu(\mc{M)}$ is the Liouville measure defined by the symplectic two-form.

Upon quantization we can write,
\al{
\label{eq:delta}
A^a_z\psi[A^a_{\bar{z}}]=\frac{2\pi}{k} \frac{\delta}{\delta A^a_{\bar{z}}} \psi[A^a_{\bar{z}}].
}

Since no currents are present, the wave-functional must satisfy the constraint $F_{z\bar{z}}\psi[A_{\bar{z}}]=0$, which is the Gauss' law of CS theory. We then make an infinitesimal gauge transformation on the wave-functional $\psi$ with parameter $\epsilon$,
\al{
\bs
\delta_\epsilon \psi[A_{\bar{z}}]=&\int d^2z\ \delta_\epsilon A^a_{\bar{z}}\ \frac{\delta\psi}{\delta A^a_{\bar{z}}} \\
=& \int d^2z\ \epsilon^a \pa{ \partial_{\bar{z}} \frac{\delta}{\delta A^a_{\bar{z}}} + i f^{abc} A^b_{\bar{z}}  \frac{\delta}{\delta A^c_{\bar{z}}} }\psi\\
=& -\frac{k}{2\pi} \int d^2z\ \epsilon^a (F^a_{z\bar{z}} - \partial_z A^a_{\bar{z}}) \psi.
\es
}
Then applying the Gauss' Law constraint $F_{z\bar{z}}\psi[A_{\bar{z}}]=0$ gives
\al{
\label{eq:infg}
\delta_\epsilon \psi[A_{\bar{z}}]= \frac{k}{2\pi} \int d^2z\ \epsilon^a  \partial_z A^a_{\bar{z}} \psi[A_{\bar{z}}].
}
This condition is solved by writing\cite{Polyakov1983121,Polyakov1984223}
\al{
\label{eq:cswf}
\psi[A_{\bar{z}}]=exp\big( kS_{WZW}(U)\big)
}
where $S_{WZW}(U)$ is the Wess-Zumino-Witten action, given by
\al{
S_{WZW}(U)=&\frac{1}{2\pi}\int \limits_{\Sigma}d^2z\ Tr\ \partial_zU \partial_{\bar{z}}U^{-1}-\frac{i}{12\pi}\int \limits_V d^3x\ \epsilon^{\mu\nu\sigma}\ Tr\ U^{-1}\partial_{\mu}UU^{-1}\partial_{\nu}UU^{-1}\partial_{\sigma}U.
}
In general, the wave-functional in \eqref{eq:cswf} can have a gauge invariant factor $\chi$ which can be found by solving the Schrodinger's equation $\mc{H}\psi=\mc{E}\psi$. But since the CS Hamiltonian for ground state is zero in the temporal gauge, we take $\chi=1$. Generally for these type of gauge theories, the wave-functional is in the form $\psi=\phi\chi$ where $\phi$ is the part that satisfies the Gauss' law constraint and $\chi$ is necessary to satisfy the Schrodinger's equation. Usually, $\chi$ is where the scale dependence is hidden. Thus, for a topological theory like CS, a constant $\chi$ is expected.

%-------------------------------------------------------------------------------------------------------------------------------------------

\subsection{The Measure}\label{ssec:csmeasure}
Using the symplectic two-form of CS theory \eqref{eq:omegacs} we can write the metric on $\mathscr{A}$, the space of gauge potentials\cite{Karabali1996135} as
\al{
\bs
ds^2_{\mathscr{A}}=&\int d^2x\ \delta A^a_i \delta A^a_i=-8\int Tr(\delta A_{\bar{z}} \delta A_z)\\
=& 8 \int Tr[D_{\bar{z}}(\delta U U^{-1})D_z(U^{\dagger -1}\delta U^{\dagger})].
\es
}
Since Cartan-Killing metric for $SL(N,\mathbb{C})$ is
\al{
ds^2_{SL(N,\mathbb{C})}=8\int Tr[(\delta U U^{-1})(U^{\dagger -1}\delta U^{\dagger})],
}
the volumes of these two spaces are related by
\al{
d\mu(\mathscr{A})=det(D_{\bar{z}}D_z)d\mu(U,U^{\dagger}).
}
This measure is not gauge invariant. To make it invariant, we may define a new matrix $H=U^{\dagger}U$ which is an element of the coset $SL(N,\mathbb{C})/SU(N)$. Now, we can write
\al{
\label{eq:CSmeasure}
d\mu(\mathscr{A})=det(D_{\bar{z}}D_z)d\mu(H).
}
The determinant is
\al{
det(D_{\bar{z}}D_z)=constant \times e^{2c_AS_{WZW}(H)}
}
where $c_A$ is the quadratic Casimir in the adjoint representation given by $c_A\delta^{ab}=f^{amn}f^{bmn}$. As shown in ref. \citen{Nair:2005iw}, the determinant is regulated using point splitting and an additional counter term was added to ensure gauge invariance. With this, the Polyakov-Wiegmann(PW)\cite{Polyakov1983121,Polyakov1984223} identity is satisfied under group action on $H$. 

Now that we have the measure and the wave functional, we can write the inner product
\al{
\langle \psi_1|\psi_2 \rangle=\int d\mu(\mathscr{A})\ e^{-K}\ \psi^*_1\psi^{ }_2.
}
Using the PW identity we get,
\al{
e^{-K}\psi^*\psi=e^{kS_{WZW}(H)}.
}
Then the inner product for $\psi$ becomes
\al{
\label{eq:csinnerproduct}
\langle \psi|\psi \rangle=\int d\mu(H)\ e^{(2c_A+k)S_{WZW}(H)}.
}
\boldmath 
\subsection*{The $ \Sigma = S^1 \times S^1 $ Case} 
\unboldmath
If the space is not simply connected, the parametrization we used in \eqref{eq:par1} needs modification. On a torus, as discussed in refs. \citen{Bos:1989kn,Nair:2005iw}, the correct parametrization is
\al{
A_{\bar{z}}=-\partial_{\bar{z}}UU^{-1}+Ui\pi(Im\ \tau)^{-1}aU^{-1},
}
where $\tau$ is the modular parameter of the torus and $a$ is a constant gauge field. The new term can be absorbed in a matrix as $V=U\ exp[i\pi(Im\ \tau)^{-1}(z-\bar{z})a]$ and then $A_{\bar{z}}$ can be once again parametrized in the form $-\partial_{\bar{z}}VV^{-1}$. But this gives rise to a new factor in the wave-functional which depends on $a$. Now the wave-functional is
\al{
\label{eq:toruspsi}
\psi[A_{\bar{z}}]=exp\big( kS_{WZW}(V)\big) \Upsilon(a).
}
Finding this new factor is not straightforward and we will not review its calculation here, but the result can be found in ref. \citen{Bos:1989kn}. 

%-------------------------------------------------------------------------------------------------------------------------------------------
%-------------------------------------------------------------------------------------------------------------------------------------------

\section{Wilson Loops in Chern-Simons Theory}\label{sec:wilcs}

\begin{figure}[h!]
\centering
\includegraphics[scale=0.43]{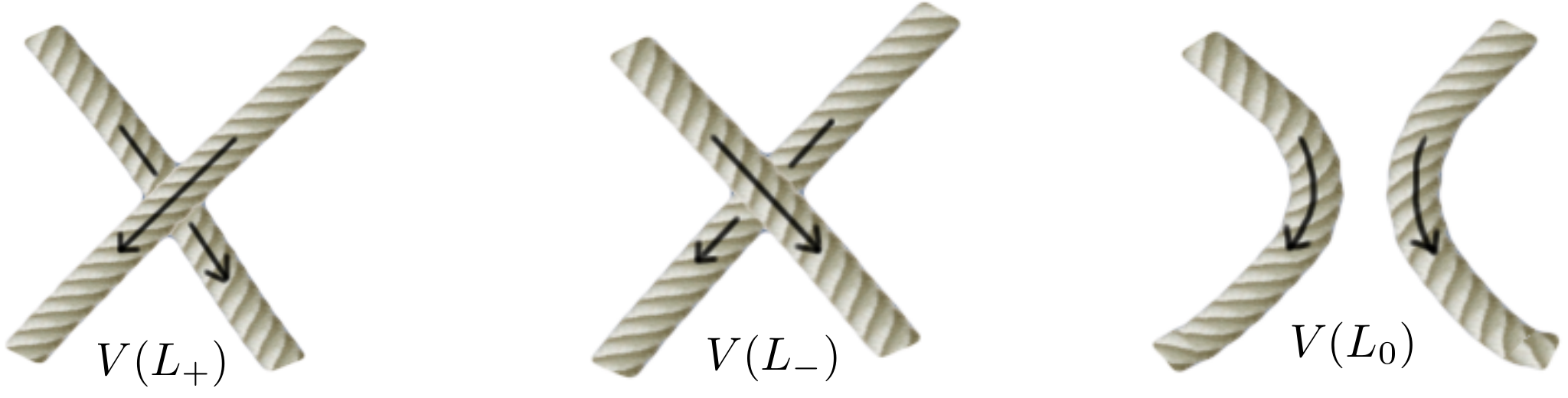}
\caption{Knot diagrams}
\label{fg:skein}
\end{figure}

The Wilson loop operator for representation $R$ and path $C$ is given by
\al{
\label{eq:wlsn}
W_R(C)=Tr_R\ \mc{P}\ e^{-\oint \limits_c A_\mu dx^\mu}.
}
As shown in ref. \citen{Witten:1988hf}, in CS theory, the expectation value of this operator can be calculated directly from skein relations without using field theory techniques. Up to some approximation, a generalized skein relation can be obtained\cite{CottaRamusino:1989rf} for WLEVs in the fundamental representation, as
\al{
\label{eq:skeinw}
\beta \langle W_{L_+} \rangle -\beta^{-1}\langle W_{L_-}\rangle=z(\beta)\langle W_{L_0}\rangle
}
where
\al{
\beta=1-i\frac{2\pi}{k}\frac{1}{2N}+\mc{O}\pa{\frac{1}{k^2}}~\text{and}~ z=-i\frac{2\pi}{k}+\mc{O}\pa{\frac{1}{k^2}}
}
and the knot diagrams are shown in \cref{fg:skein}.

In the temporal gauge with complex coordinates and in representation $R$, \eqref{eq:wlsn} becomes
\al{
W_R(C)=Tr_R\ \mc{P}\ e^{-\oint \limits_c (A_zdz+A_{\bar{z}}d\bar{z})}.
}
Since $A_z$ is the derivative with respect to $A_{\bar{z}}$ and it acts on everything on its right, expanding this path ordered exponential leads to a very difficult calculation. To avoid this, instead of using the usual definition of the Wilson loop, we would like to use a \emph{Wilson loop-like} observable defined as
\al{
\mc{W}_R(C)=Tr_R\ \mc{P}\ e^{-\oint \limits_c (\mc{A}_zdz+A_{\bar{z}}d\bar{z})}=Tr_R\ U(x,x,C),
}
where $\mc{A}_z$ is given by $\partial_z A_{\bar{z}}-\partial_{\bar{z}}\mc{A}_z+[\mc{A}_z,A_{\bar{z}}]=0$. Since $F_{z\bar{z}}=0$ from the Gauss' law, replacing $A_z$ with $\mc{A}_z$ does not change the general properties of the Wilson loop when evaluated on states that live on the constraint hyper-surface. But $U$ is defined to be path independent, so it seems like skein relations are trivially satisfied. However, this is not true since the path independence is only on $\Sigma$, because we are forcing flatness only on the $z\bar{z}$ component of the curvature. So, one is allowed to make small deformations in the time direction, piercing $\Sigma$ to get skein relations.

In the previous section, we have shown that the theory is given by the action $S_{WZW}(H)$, thus we can use gauge invariant WZW currents $J_{\bar{z}}=-\partial_{\bar{z}}HH^{-1}$ and $J_z=H^{-1}\partial_z H$ to write gauge invariant observables similar to Wilson loops\cite{Karabali:1998yq}. The gauge fields in $\mc{W}_R(C)$ can be written as $SL(N,\mathbb{C})$ transformed WZW currents,
\al{
\label{eq:AJ}
\bs
A_{\bar{z}}=&-\partial_{\bar{z}}UU^{-1}=U^{\dagger-1}J_{\bar{z}}U^{\dagger}+U^{\dagger-1}\partial_{\bar{z}}U^{\dagger},\\
\mc{A}_z=&-\partial_zUU^{-1}=U^{\dagger-1}J_zU^{\dagger}+U^{\dagger-1}\partial_zU^{\dagger}.
\es
}
Thus, we can write $\mc{W}$ in terms of H
\al{
\mc{W}_R(C,H)=Tr_R\ \mc{P}\ e^{\ \oint \limits_c (\partial_zHH^{-1}dz+\partial_{\bar{z}}HH^{-1}d\bar{z})}.
}
The expectation value of $\mc{W}$ is given by
\al{
\langle \mc{W}_R(C) \rangle=\int d\mu(H)\ e^{(2c_a+k)S_{WZW}(H)}\mc{W}_R(C,H).
}

%-------------------------------------------------------------------------------------------------------------------------------------------
%-------------------------------------------------------------------------------------------------------------------------------------------

\section{Topologically Massive Yang-Mills Theory}\label{sec:tmym}

TMYM action is given by
\al{
\label{eq:actiontmym}
\bs
S_{TMYM}=&S_{CS}+S_{YM}\\
=&-\frac{k}{4\pi}\int \limits_{\Sigma \times[t_i,t_f]} {d^3x }\ \epsilon^{\mu\nu\alpha}\ Tr \pa{A_\mu \partial_{\nu} A_{\alpha} + \frac{2}{3}A_\mu A_\nu A_\alpha}\\
&-\frac{k}{4\pi}\frac{1}{4m}\int \limits_{\Sigma\times[t_i,t_f]} {d^3x }\ Tr\  F_{\mu\nu}F^{\mu\nu}.
\es
} 
Here $m$ is called the topological mass. Our definition of topological mass differs by a factor $\frac{k}{4\pi}$ compared to the literature. We made this choice so that studying different values of $k$ does not change the balance of the theory in either pure YM or pure CS direction. That is decided only by the value of $m$. With this choice of constants, the equations of motion are $k$ free, as
\al{
\epsilon^{\mu\alpha\beta}F_{\alpha\beta}+\frac{1}{m} D_{\nu}F^{\mu\nu}=0
}
where $D_\mu \bullet=\partial_\mu \bullet + [A_\mu,\bullet]$.
To simplify the notation, we define
\al{
\label{eq:tilde0}
\tilde A_{\mu}=A_{\mu}+\frac{1}{2m}\epsilon_{\mu\alpha\beta}F^{\alpha\beta}.
}
Here, $\tilde{A}$ is not a field redefinition. It is just a shorthand notation to make equations easy to compare with pure CS theory. 
From \eqref{eq:tilde0}, it can be seen that $\tilde A_\mu$ transforms like a gauge field since $F^{\alpha\beta}$ is gauge covariant. For future convenience, using $\tilde A_\mu$ as a connection, we define a new covariant derivative $\tilde D_\mu \bullet=\partial_\mu \bullet + [\tilde A_\mu,\bullet]$.

Using complex coordinates and temporal gauge, the conjugate momenta are given by $\tilde{A}$,
\al{
\Pi^{a z}=\frac{ik}{4\pi} \tilde A^a_{\bar{z}}~\ \text{and}~\ \Pi^{a \bar{z}}=-\frac{ik}{4\pi}  \tilde A^a_z,
}
with
\al{
\label{eq:Az}
\tilde A_{z}=A_{z}+E_{z}~\ \text{and}~\ \tilde A_{\bar{z}}=A_{\bar{z}}+E_{\bar{z}}
}
and where
\al{
\label{eq:E}
E_{z}=\frac{i}{2m}F^{0\bar{z}} \ ~\text{and}~\  E_{\bar{z}}=-\frac{i}{2m}F^{0z}.
}
The conjugate momenta of TMYM theory transform like gauge fields and this feature gives the theory a CS-like behavior in some sense. Thus, we will be able to borrow many of the features of CS theory in the following analysis.

The symplectic two-form for this theory is
\al{
\bs
\label{eq:omegatmym}
\Omega=&\frac{ik}{4\pi}\int \limits_{\Sigma}(\delta \tilde A^a_{\bar{z}} \delta A^a_z+\delta A^a_{\bar{z}}\delta \tilde A^a_z )\\
=&\frac{ik}{4\pi}\int \limits_{\Sigma}(2\delta A^a_{\bar{z}} \delta A^a_z +\delta E^a_{\bar{z}} \delta A^a_z+ \delta A^a_{\bar{z}}\delta E^a_z  ).
\es
}
From this equation, it can be seen that the TMYM phase space consists of two CS-like parts. This can be seen more clearly under a coordinate transformation. Instead of using $A_z$ and $A_{\bar{z}}$, we could use $B_z=\frac{1}{2}(A_1+i\tilde A_2)$, $C_z=\frac{1}{2}(\tilde A_1+i A_2)$ and their complex conjugates. This would allow us to write $\Omega$ in the form $\delta B_z \delta B_{\bar{z}} + \delta C_z \delta C_{\bar{z}}$. Thus, the phase space of TMYM theory can be written as two CS phase spaces.
%-------------------------------------------------------------------------------------------------------------------------------------------

\subsection{The Wave-Functional}

We choose the K\"ahler polarization that makes $\psi$ only $A_{\bar{z}}$ and $\tilde{A}_{\bar{z}}$ dependent. The pre-quantum wave-functional is $\Phi[A_z,A_{\bar{z}},\tilde A_z, \tilde A_{\bar{z}}]=e^{-\frac{1}{2}K}\psi[A_{\bar{z}},\tilde A_{\bar{z}}]$, where $K=\frac{k}{4\pi}\int_\Sigma (\tilde A^a_{\bar{z}} A^a_z+ A^a_{\bar{z}} \tilde A^a_z )$ is the K\"ahler potential. Now, we can write
\al{
\label{eq:deltatmym}
A^a_z\psi=\frac{4\pi}{k} \frac{\delta}{\delta \tilde A^a_{\bar{z}}} \psi\ ~\text{and}\ ~\tilde A^a_z\psi=\frac{4\pi}{k} \frac{\delta}{\delta A^a_{\bar{z}}}\psi .
}
We make an infinitesimal gauge transformation on $\psi$ as
\al{
\label{eq:infgauge}
\delta_\epsilon \psi[A_{\bar{z}},\tilde{A}_{\bar{z}}]=\int d^2z\ \pa{ \delta_\epsilon A^a_{\bar{z}} \frac{\delta\psi}{\delta A^a_{\bar{z}}}  +\ \delta_\epsilon \tilde{A}^a_{\bar{z}} \frac{\delta\psi}{\delta \tilde{A}^a_{\bar{z}}}}.
}
Using \eqr{eq:deltatmym} and $\delta A^a_{\bar{z}}=D_{\bar{z}}\epsilon^a$, $\delta \tilde{A}^a_{\bar{z}}=\tilde{D}_{\bar{z}}\epsilon^a$, we get
\al{
\bs
\delta_\epsilon \psi=& \int d^2z\ \epsilon^a \pa{\tilde{D}_{\bar{z}}\frac{\delta}{\delta \tilde{A}^a_{\bar{z}}} + D_{\bar{z}}\frac{\delta}{\delta A^a_{\bar{z}}} } \psi\\
=&\frac{k}{4\pi} \int d^2z\ \epsilon^a \pa { \partial_z \tilde{A}^a_{\bar{z}}+ \partial_z A^a_{\bar{z}} -2F_{z\bar{z}}- D_z E_{\bar{z}}+D_{\bar{z}}E_z  }\psi
\es
}
The generator of infinitesimal gauge transformations for this theory is
\al{
\label{eq:gausstmym}
G^a=\frac{ik}{4\pi}(2F_{z\bar{z}} + D_z E_{\bar{z}} - D_{\bar{z}}E_z)
}
while $G^a=\frac{ik}{2\pi}F_{z\bar{z}}$ being the generator for the pure CS theory as $E$-fields go to zero.
After applying the Gauss' law $G^a\psi=0$, $\delta_\epsilon \psi$ becomes
\al{
\label{eq:infg2}
\delta_\epsilon \psi= \frac{k}{4\pi} \int d^2z\ \epsilon^a \pa{\partial_z \tilde A^a_{\bar{z}}+\partial_z A^a_{\bar{z}} } \psi.
}
This equation is similar to \eqr{eq:infg}. As they transform identically, $\tilde A$ can be parametrized the same way as $A$, using a different SL(N,$\mathbb{C}$) matrix $\tilde U$,
\al{
\tilde A_{\bar z}=-\partial_{\bar z}\tilde U \tilde U^{-1}~\text{and}~\tilde A_z=\tilde U^{{\dagger}{-1}}\partial_z \tilde U^{\dagger}.
}
The solution for the condition \eqr{eq:infg2} is
\al{
\label{eq:wftmym}
\psi[A_{\bar{z}},\tilde{A}_{\bar{z}}]=exp\pb{\frac{k}{2}\big(S_{WZW}(\tilde U)+S_{WZW}( U)\big)}\chi
}
where $\chi$ is the gauge invariant part of $\psi$ that is required to satisfy the Schrodinger's equation. Notice that \eqref{eq:wftmym}
reduces to CS wave-functional \eqr{eq:cswf} as expected, when topological mass approaches infinity, which is equivalent to dropping the tilde symbol. $\chi$ should be equal to one in this limit.

To understand the role of the new $\tilde{U}$ matrix, we can relate it to $U$ by rewriting \eqr{eq:Az} as
\al{
\label{eq:tilde}
\partial_{\bar z}\tilde U \tilde U^{-1}=\partial_{\bar z} U  U^{-1} + \frac{i}{2m}F^{0z}.
}
It turns out that $\tilde{U}$ is well behaved and solvable. Using the assumption $\tilde U=UM$, we can solve \eqref{eq:tilde} for $M$, viz;
\al{
\label{eq:M}
M(z,\bar{z})=\mc{P}exp \pa{ \frac{i}{2m} \int_0^{\bar{z}} \mc{F}^{0w}d\bar{w} }.
}
Here $\mc{F}^{0z}=U^{-1}F^{0z}U$ and it is gauge invariant.  

With this new gauge invariant matrix $M$, the electric field components can be written as
\al{
\label{eq:EU}
E_z=U^{\dagger -1}M^{\dagger -1}\partial_z M^\dagger U^\dagger \ ~\text{and}~\ E_{\bar{z}}=-U\partial_{\bar{z}} M M^{-1} U^{-1}.
}

%-------------------------------------------------------------------------------------------------------------------------------------------

\subsubsection*{The Hamiltonian}\label{sec:hamilt}

With no charges present, the Hamiltonian gets no contribution from the CS term in the temporal gauge. With $\alpha=\frac{4\pi}{k}$, $B=\frac{ik}{2\pi}F^{z\bar{z}}$ and using Euclidean metric, the Hamiltonian is
\al{
\mc{H}=\frac{m}{2\alpha}(E^a_{\bar{z}} E^a_z + E^a_z E^a_{\bar{z}})+\frac{\alpha}{m} B^a B^a.
}
Using \eqref{eq:deltatmym} we get
\al{
\bs
E^a_z(x)E^b_{\bar{z}}(x')\psi =&\ \alpha \pa{\frac{\delta}{\delta A^a_{\bar{z}}(x)} - \frac{\delta}{\delta \tilde{A}^a_{\bar{z}}(x)}}(\tilde A^b_{\bar{z}}(x')-A^b_{\bar{z}}(x'))\psi, 
\es
}
which gives the commutator
\al{
\label{eq:ecom}
[E^a_z(x),E^b_{\bar{z}}(x')]=-2 \alpha\ \delta^{ab}\delta^{(2)}(x-x').
}
Here, $E^a_z$ can be interpreted as an annihilation operator and $E^b_{\bar{z}}$ as a creation operator \cite{Grignani1997360}.
To get rid of the infinity, the Hamiltonian has to be normal ordered as
\al{
\mc{H}=\frac{m}{\alpha}E^a_{\bar{z}} E^a_z +\frac{\alpha}{m} B^a B^a.
}
To simplify the notation we define
\al{
\phi=exp\pb{\frac{k}{2}\big(S_{WZW}(\tilde U)+S_{WZW}( U)\big)},
}
hence $\psi=\phi \chi$. Derivatives of $\phi$ give the gauge field $\mc{A}$ defined in \eqref{eq:scriptA} and its tilde version\cite{Nair:2005iw},
\al{
\label{eq:phi}
\tilde A^a_z\phi=\frac{4\pi}{k}\frac{\delta \phi}{\delta A^a_{\bar{z}}}=\mc{A}^a_z \phi\  ~\text{and}~\  A^a_z\phi=\frac{4\pi}{k}\frac{\delta \phi}{\delta \tilde{A}^a_{\bar{z}}}=\tilde{\mc{A}}^a_z \phi.
}
The holomorphic component of the $E$-field acting on $\psi$ is
\al{
E^a_z\psi=&\frac{4\pi}{k}\pa{\frac{\delta \phi}{\delta A^a_{\bar{z}}}-\frac{\delta \phi}{\delta \tilde{A}^a_{\bar{z}}}}\chi+\frac{4\pi}{k}\pa{\frac{\delta \chi}{\delta A^a_{\bar{z}}}-\frac{\delta \chi}{\delta \tilde{A}^a_{\bar{z}}}}\phi.
}
With defining $\mc{E}_z=\tilde{\mc{A}}_z - \mc{A}_z=-U\partial_z M M^{-1} U^{-1}$, we can write
\al{
\label{eq:EE}
E^a_z\psi=-\mc{E}^a_z\psi+\frac{4\pi}{k}\pa{\frac{\delta \chi}{\delta A^a_{\bar{z}}}-\frac{\delta \chi}{\delta \tilde{A}^a_{\bar{z}}}}\phi.
}
The magnetic field acting on $\psi$ is
\al{
\label{eq:B}
\bs
F^a_{z\bar{z}}\psi=&(\partial_z A^a_{\bar{z}}-D_{\bar{z}}A^a_z)\psi\\
=&D_{\bar{z}}(\mc{A}^a_z-A^a_z)\psi\\
=&D_{\bar{z}}\pa{-\mc{E}^a_z\psi-\frac{4\pi}{k}\frac{\delta \chi}{\delta \tilde{A}^a_{\bar{z}}}\phi}.
\es
}
The vacuum wave-functional is given by $\mc{H}\psi_0=0$, or
\al{
\label{eq:Hamilt}
E^a_{\bar{z}}E^a_z \psi_0 +\frac{1}{64m^2} F^a_{z\bar{z}} F^a_{z\bar{z}} \psi_0 =0.
}
The first term in \eqref{eq:Hamilt} is second order in $1/m$, while the Gauss' law forces the second term to be fourth order. We will continue our analysis with finite large values of $m$ where the potential energy term is negligible. Then, $E^a_z$ annihilates the vacuum, $E^a_z\psi_0=0$. This condition is solved by writing
\al{
\label{eq:chi0}
\bs
\chi_0^{ }=&exp \pa{-\frac{k}{8\pi}\int \limits_{\Sigma} (\tilde{A}^a_{\bar{z}}-A^a_{\bar{z}})\mc{E}^a_z}=exp \pa{-\frac{k}{8\pi}\int \limits_{\Sigma}E^a_{\bar{z}}\mc{E}^a_z}.
\es
}
This solution is gauge invariant as required. In terms of $SL(N,\mathbb{C})$ matrices, it is a function of just the gauge invariant matrix $M$, which was defined in \eqref{eq:M},
\al{
\chi_0^{ }=exp \pa{\frac{k}{4\pi}\int \limits_{\Sigma}Tr (E_{\bar{z}}\mc{E}_z)}=exp \pa{-\frac{k}{4\pi}\int \limits_{\Sigma}Tr (\partial_{\bar{z}}MM^{-1}\partial_zMM^{-1})}.
}

%-------------------------------------------------------------------------------------------------------------------------------------------

\subsection{The Measure}

Using the symplectic two-form \eqref{eq:omegatmym} we write the metric
\al{
\bs
ds^2_{\mathscr{A}}=&-4\int Tr(\delta \tilde{A}_{\bar{z}} \delta A_z+\delta A_{\bar{z}}\delta \tilde{A}_z)\\
=&\ 4 \int Tr[\tilde{D}_{\bar{z}}(\delta \tilde{U} \tilde{U}^{-1})D_z(U^{\dagger -1}\delta U^{\dagger})+D_{\bar{z}}(\delta U U^{-1})\tilde{D}_z(\tilde{U}^{\dagger -1}\delta \tilde{U}^{\dagger})].
\es
}
Similar to the analysis in \autoref{ssec:csmeasure}, the gauge invariant measure for this case is
\al{
\label{eq:meas}
d\mu(\mathscr{A})=det(\tilde{D}_{\bar{z}}D_z) det(D_{\bar{z}}\tilde{D}_z)d\mu(\tilde{U}^{\dagger}U)d\mu(U^{\dagger}\tilde{U})
}
where, for a certain choice of local counter terms ($\int Tr(\tilde A_{\bar{z}}A_z+\tilde A_z A_{\bar{z}})$),
\al{
det(\tilde{D}_{\bar{z}}D_z) det(D_{\bar{z}}\tilde{D}_z)=constant \times e^{2c_A\big(S_{WZW}(\tilde{U}^{\dagger}U)+S_{WZW}(U^{\dagger}\tilde{U})\big)}.
}
To simplify the notation we define a new matrix $N=\tilde{U}^{\dagger}U$. Since $\tilde{U}$ transforms like $U$, $N$ is gauge invariant. With this definition the measure becomes
\al{
\label{eq:measure}
d\mu(\mathscr{A})=constant \times e^{2c_A\big(S_{WZW}(N)+S_{WZW}(N^{\dagger})\big)} d\mu(N)d\mu(N^{\dagger}).
}
To find the inner product, using PW identity we write
\al{
e^{-K_{TMYM}}\psi_{TMYM}^*\psi^{ }_{TMYM}=e^{\frac{k}{2}\big(S_{WZW}(N)+S_{WZW}(N^{\dagger})\big)}\chi_0^*\chi^{ }_0
}
and from \eqref{eq:chi0} $\chi_0^*\chi^{ }_0$ (for large $m$) is 
\al{
\chi_0^*\chi^{ }_0=exp \pa{-\frac{k}{8\pi}\int \limits_{\Sigma}(E^a_z \mc{E}^a_{\bar{z}}+E^a_{\bar{z}}\mc{E}^a_z )}.
}
Then the inner product for the vacuum state is
\al{
\label{eq:psipsi0}
\langle \psi_0|\psi_0\rangle=\int d\mu(N)d\mu(N^{\dagger})\ e^{(2c_A+\frac{k}{2})\big(S_{WZW}(N)+S_{WZW}(N^{\dagger})\big)}e^{-\frac{k}{8\pi}\int(E^a_z \mc{E}^a_{\bar{z}}+E^a_{\bar{z}}\mc{E}^a_z )}.
}
Since $\chi_0^*\chi^{ }_0=1+\mc{O}(1/m^2)$, we can neglect the second and higher order contributions at large scales compared to $1/m$, which leads to an almost topological theory in the near CS limit as
\al{
\label{eq:psipsi}
\langle \psi_0|\psi_0\rangle_{TMYM_k}{\approx} \int d\mu(N)d\mu(N^{\dagger})\ e^{(2c_A+\frac{k}{2})\big(S_{WZW}(N)+S_{WZW}(N^{\dagger})\big)}=\langle\psi|\psi\rangle ^2_{CS_{k/2}}.
}
Here the label ${TMYM_k}$ means that the inner product is calculated for TMYM theory with CS level number $k$. Similarly, $CS_{k/2}$ means the inner product is calculated for pure CS theory with level $k/2$. On the pure CS side, it is important to make this replacement of the level number to make the equivalence work. These two half-CS theories are not separately gauge invariant for odd values of $k$, but the sum of the two is. Each piece transforms as $\frac{1}{2}S_{CS}(A^g)\rightarrow\frac{1}{2}S_{CS}(A)+\pi k \omega(g)$ where $\omega(g)$ is the winding number. Then, the sum of the two will bring an extra $2\pi k \omega(g)$ that will not change the value of the path integral, even for odd values of $k$. In other words, any integer value of $k$ is sufficient to make the left hand side of \eqref{eq:psipsi} gauge invariant. But if one wants to make the two split CS parts separately gauge invariant on the right hand side of \eqref{eq:psipsi}, even values of $k$ must be chosen. 

This equivalence at large distances $(d>1/m)$ comes from the fact that the phase space of TMYM theory is a direct sum of two CS-like phase spaces. Thus, the classical equivalence discussed in refs. \citen{Lemes:1997vx, Lemes:1998md, Quadri:2002ni} does not work at the quantum level as it is, because of different phase space dimensionality of two theories.

In \eqref{eq:meas}, gauge invariance can be obtained in a different way by choosing other counter terms.
Choosing $\int Tr(A_{\bar{z}}A_z+\tilde A_z \tilde A_{\bar{z}})$ leads to
\al{
d\mu(\mathscr{A})=constant \times e^{2c_A\big(S_{WZW}(H)+S_{WZW}(\tilde H)\big)} d\mu(H)d\mu(\tilde H).
}
With this option $e^{-K}\psi^*\psi$ part differs by $e^{-\int E^2}=1+\mc{O}(1/m^2)$ and the inner product can still be written in the form
\al{
\label{eq:H-Htilde}
\langle \psi_0|\psi_0\rangle_{TMYM_k}{\approx} \int d\mu(H)d\mu(\tilde H)\ e^{(2c_A+\frac{k}{2})\big(S_{WZW}(H)+S_{WZW}(\tilde H)\big)}=\langle\psi|\psi\rangle ^2_{CS_{k/2}}.
}
Thus, the CS splitting can still be observed in the near CS limit. Just like $N$ and $N^\dagger$, $H$ and $\tilde H$ are elements of $SL(N,\mathbb{C})/SU(N)$. However, the $N$, $N^\dagger$ choice seems to be more \emph{natural} than $H$, $\tilde H$ because tilde and non-tilde variables are mixed in \eqref{eq:omegatmym}.

%-------------------------------------------------------------------------------------------------------------------------------------------
%-------------------------------------------------------------------------------------------------------------------------------------------

\section{Topologically Massive Gravity}\label{sec:TMG}

There is an analogous CS splitting in topologically massive AdS gravity that can be seen clearly in refs. \citen{witten1988,Carlip2008272, Carlip2}. For a dynamical metric $\gamma_{\mu\nu}$, this model has the action
\al{
\label{eq:tmgS}
S=\int d^3x \left[ -\sqrt{-\gamma}(R-2\Lambda)+\frac{1}{2\mu}\epsilon^{\mu\nu\rho}\left(  \Gamma^\alpha_{\mu\beta}\partial_\nu \Gamma^\beta_{\rho\alpha} +\frac{2}{3}\Gamma^\alpha_{\mu\gamma} \Gamma^\gamma_{\nu\beta} \Gamma^\beta_{\rho\alpha}    \right)  \right].
}
With defining
\al{
A^{\pm}{}_{\mu}{}^a{}_b[e]=\omega_{\mu}{}^a{}_b[e] \pm \epsilon^a{}_{bc} e_\mu{}^c
}
where $e_\mu{}^a$ is the dreibein and $\omega_{\mu}{}^a{}_b[e]$ is the torsion-free spin connection, the action \eqref{eq:tmgS} can be written as
\al{
S[e]=-\frac{1}{2}\pa{1-\frac{1}{\mu}}S_{CS}\big[A^+[e]\big]+\frac{1}{2}\pa{1+\frac{1}{\mu}}S_{CS}\big[A^-[e]\big]
}
where 
\al{
S_{CS}[A]=\frac{1}{2}\int \epsilon^{\mu\nu\rho}\pa{A_\mu{}^a{}_b \partial_\nu A_\rho{}^b{}_a + \frac{2}{3} A_\mu{}^a{}_c A_\nu{}^c{}_b A_\rho{}^b{}_a}.
}
For our interests, the main difference between this gravity model and TMYM theory is that the latter has a mass gap, therefore it has a CS like behavior only at large distances. Thus, TMYM theory is analogous to this model only in the near CS limit. In this analogy, small $\mu$ corresponds to large $m$. In the near CS limit, the gravity theory splits into two CS parts that are added together, each with half the level, similar to what we have observed for TMYM theory.

%-------------------------------------------------------------------------------------------------------------------------------------------
%-------------------------------------------------------------------------------------------------------------------------------------------

\section{Wilson Loops in Topologically Massive Yang-Mills Theory}\label{sec:wilson}

With the new gauge field $\tilde{A}$, we would like to define a new loop operator
\al{
\label{eq:tildewlsn}
T_R(C)=Tr_R\ \mc{P}\ e^{-\oint \limits_c \tilde{A}_\mu dx^\mu}.
}
Just as the traditional Wilson loop, this operator is gauge invariant and it is an observable of the theory. To make a physical interpretation of this loop, we will check to see if it satisfies a 't Hooft-like algebra with the Wilson loop. To simplify the calculation, we will look at the abelian case with the following loops that live on $\Sigma$
\al{
W(C)=e^{i\oint \limits_c (A_z dz+A_{\bar{z}} d\bar{z})}\ ~ \text{and}\ ~ T(C)=e^{i\oint \limits_c (\tilde{A}_z dz+\tilde{A}_{\bar{z}} d\bar{z})}.
}
For the abelian case, the canonical relations differ from \eqref{eq:deltatmym} by a factor of 2. Then two loop operators satisfy a 't Hooft-like algebra
\al{
\label{eq:thooft}
T(C_1)W(C_2)=e^{\frac{2\pi i}{k}l(C_1,C_2)}W(C_2)T(C_1),
}
where $l(C_1,C_2)$ is the intersection number of $C_1$ and $C_2$ , which can only take values $0,\pm 1$. We cannot get a Dirac-like quantization condition since $k$ appears in the denominator and we want it to be a large integer to make the skein relations work. Therefore, the only option to make these operators commute is to have $l(C_1,C_2)=0$ thus, two loops cannot share a point. Equation \eqref{eq:thooft} lets us to interpret $T(C)$ as a 't Hooft-like loop for TMYM theory.

Working with the holomorphic polarization leads to the same problem we had in the pure CS case. $A_z$ and $\tilde{A}_z$ are derivatives with respect to $\tilde{A}_{\bar{z}}$ and $A_{\bar{z}}$. This makes the path ordered exponential very complicated. To avoid this problem, we use a similar technique: Instead of using the traditional Wilson loop, we will calculate the expectation value of two loop operators that we define by $Tr\ U(x,x,C)$ and $Tr\ \tilde{U}(x,x,C)$ or
\al{
\mc{W}_R(C)=Tr_R\ \mc{P}\ e^{-\oint \limits_c (\mc{A}_zdz+A_{\bar{z}}d\bar{z})}\ ~
\text{and}~ \ 
\mc{T}_R(C)=Tr_R\ \mc{P}\ e^{-\oint \limits_c (\tilde{\mc{A}}_zdz+\tilde{A}_{\bar{z}}d\bar{z})}.
}
Once again these can be written in terms of WZW currents $-\partial_{\bar{z}}NN^{-1}$,  $-\partial_zNN^{-1}$,  $-\partial_{\bar{z}}N^{\dagger}N^{\dagger-1}$ and $-\partial_zN^{\dagger}N^{\dagger-1}$ as
\al{
\mc{W}_R(C)=Tr_R\ \mc{P}\ e^{\ \oint \limits_c (\partial_zNN^{-1}dz+\partial_{\bar{z}}NN^{-1}d\bar{z})}\ ~
\text{and}~ \ 
\mc{T}_R(C)=Tr_R\ \mc{P}\ e^{\ \oint \limits_c (\partial_zN^{\dagger}N^{\dagger-1}dz+\partial_{\bar{z}}N^{\dagger}N^{\dagger-1}d\bar{z})}.
}
These WZW currents are $SL(N,\mathbb{C})$ transformed $A$ and $\tilde A$ fields, just like in \eqref{eq:AJ}.

There is an interesting expectation value that we can calculate;
\al{
\bs
\langle \mc{W}_{R_1}(C_1)\mc{T}_{R_2}(C_2) \rangle=&\int d\mu(\mathscr{A}) \psi_0^*\mc{W}_{R_1}(C_1)\mc{T}_{R_2}(C_2)\psi_0\\
=&\int d\mu(N)d\mu(N^{\dagger})\ e^{(2c_A+\frac{k}{2})\big(S_{WZW}(N)+S_{WZW}(N^{\dagger})\big)}\ \mc{W}_{R_1}(C_1,N)\mc{T}_{R_2}(C_2,N^{\dagger})\\
&+\mc{O}(1/m^2).
\es
}
This leads to an equivalence between the observables of CS and TMYM theory in the near CS limit. $\mc{W}_R(C)$ being only $N$ dependent and $\mc{T}_R(C)$ being only $N^{\dagger}$ dependent lets us to write
\begin{subequations}
\label{eq:equivalence}
\al{
\label{eq:equivalence1}
\langle \mc{W}_R(C)\rangle_{TMYM_{2k}} = \langle \mc{W}_R(C)\rangle_{CS_{k}}+\mc{O}(1/m^2),
}
\al{
\label{eq:equivalence2}
\langle \mc{T}_R(C)\rangle_{TMYM_{2k}} = \langle \mc{W}_R(C)\rangle_{CS_{k}}+\mc{O}(1/m^2)
}
and
\al{
\label{eq:equivalence3}
\langle \mc{W}_{R_1}(C_1)\mc{T}_{R_2}(C_2)\rangle_{TMYM_{2k}} = \bigg(\langle \mc{W}_{R_1}(C_1)\rangle_{CS_{k}}\bigg)\bigg(\langle \mc{W}_{R_2}(C_2)\rangle_{CS_{k}}\bigg)+\mc{O}(1/m^2).
}
\end{subequations}
To generalize these for $n$ loops, we can write
\begin{subequations}
\label{eq:equivalencegnrl}
\al{
\label{eq:equivalence1gnrl}
\langle \mc{W}_{R_1}(C_1)\textellipsis&\mc{W}_{R_n}(C_n)\rangle_{TMYM_{2k}} = \langle \mc{W}_{R_1}(C_1)\textellipsis\mc{W}_{R_n}(C_n))\rangle_{CS_{k}}+\mc{O}(1/m^2),
}
\al{
\label{eq:equivalence2gnrl}
\langle \mc{T}_{R_1}(C_1)\textellipsis&\mc{T}_{R_n}(C_n)\rangle_{TMYM_{2k}} = \langle\mc{W}_{R_1}(C_1)\textellipsis\mc{W}_{R_n}(C_n))\rangle_{CS_{k}}+\mc{O}(1/m^2)
}
and for mixed $n$ Wilson-like and $m$ 't Hooft-like loops,
\al{
\label{eq:equivalence3gnrl}
\bs
\langle \mc{W}_{R_1}(C_1)\textellipsis&\mc{W}_{R_n}(C_n)\mc{T}_{R'_1}(C'_1)\textellipsis\mc{T}_{R'_m}(C'_m)\rangle_{TMYM_{2k}}\\
& = \bigg(\langle \mc{W}_{R_1}(C_1)\textellipsis\mc{W}_{R_n}(C_n)\rangle_{CS_{k}}\bigg)\bigg(\langle\mc{W}_{R'_1}(C'_1)\textellipsis\mc{W}_{R'_m}(C'_m)\rangle_{CS_{k}}\bigg)\\
&+\mc{O}(1/m^2).
\es
}
\end{subequations}

Although \eqref{eq:psipsi} is gauge invariant even for odd values of $k$ as we have explained before, writing WLEVs of TMYM theory in terms of WLEVs of CS theory requires the two split CS parts to be separately gauge invariant. For this reason, we have used even level numbers on the left hand side of \eqref{eq:equivalence} and \eqref{eq:equivalencegnrl}. Notice that this gauge invariance issue arose only because we wanted to arrive to an equivalence between the observables of TMYM and CS theories. Otherwise, $\langle  \mc{W}_{R_1}(C_1)\mc{T}_{R_2}(C_2)\rangle_{TMYM_k}$ is gauge invariant in its own right for all integer values of $k$.

It seems like the set of equivalences we obtained also work for the case where $\Sigma=S^1\times S^1$. On a torus, similar to \eqref{eq:toruspsi} TMYM wave-functional becomes
\al{
\psi[A_{\bar{z}},\tilde{A}_{\bar{z}}]=exp\pb{\frac{k}{2}(S_{WZW}(\tilde V)+S_{WZW}( V))}\Upsilon(\tilde{a})\Upsilon(a)\chi.
}
On the TMYM side, one needs to integrate over $V,\tilde{V},a,\tilde{a}$ and on the CS side only over $V$ and $a$. Although it requires a more careful analysis, \eqref{eq:equivalencegnrl} seem to work on a torus as well. In principle, there is no reason to expect it to not work on any orientable $\Sigma$.

%-------------------------------------------------------------------------------------------------------------------------------------------
%-------------------------------------------------------------------------------------------------------------------------------------------

\section{Conclusion}\label{sec:conc}

We have shown that due to the existence of a mass gap, TMYM theory in the \emph{near} CS limit is an almost topological field theory that consists of two copies of CS, similar to the topologically massive AdS gravity model. One copy is associated with the matrix $N$ and the other with $N^{\dagger}$, each with half the level of the original CS term in the TMYM Lagrangian. Separately momentum and position Hilbert spaces of TMYM theory can be thought of CS Hilbert spaces with half the level. In the $m\rightarrow\infty$ limit, where $N=N^{\dagger}=H$, these two CS theories add up to give one CS with the original level number $k$, as
\al{
e^{\frac{k}{2}\big(S_{WZW}(N)+S_{WZW}(N^{\dagger})\big)} \overset{m\rightarrow\infty}\longrightarrow e^{kS_{WZW}(H)}.
}
Although the integrand behaves well as $m\rightarrow\infty$, this limit is delicate for the integral measure. Studying large values of $m$ does not cause any problem, but taking it to infinity reduces the phase space dimension from four to two, thus a change in the integral measure becomes necessary. Except for this phase space reduction, dropping the tilde symbol gives the correct CS limit in our calculations. In this limit, the metric of space of gauge potentials reduces as
\al{
\label{eq:reduction}
ds^2_{\mathscr{A}}=-4\int Tr(\delta \tilde{A}_{\bar{z}} \delta A_z+\delta A_{\bar{z}}\delta \tilde{A}_z)
\overset{m\rightarrow\infty}\longrightarrow ds^2_{\mathscr{A}}=-8\int Tr(\delta A_{\bar{z}} \delta A_z).
}
For the left hand side of \eqref{eq:reduction}, the measure is given by \eqref{eq:measure} while for the right hand side, it is given by \eqref{eq:CSmeasure}. Thus, the measure needs to be replaced with \eqref{eq:CSmeasure} in the pure CS limit. Although this reduction occurs beautifully in the metric, the volume \eqref{eq:measure} does not automatically reduce to \eqref{eq:CSmeasure} in our notation. Not switching to the correct volume element results in duplicate integration over $H$, since in the pure CS limit $N=N^{\dagger}=H$. This comes from the fact that the phase space of TMYM theory consists of two CS phase spaces.

A different problem exists for the pure YM limit $m\rightarrow0$. In this case, dimensionality of the phase space does not change, but since $E$-fields do not gauge transform like $\tilde{A}$ fields, our parametrization and measure do not work. But the main problem with studying the pure YM limit comes from not knowing the magnetic field contribution in the wave-functional, which becomes the dominant part in this limit. To get the magnetic field contribution, \eqref{eq:Hamilt} needs to be fully solved without using the strong coupling limit.

Both of the limits we discussed above cause some problems. It seems that transition from pure CS to TMYM or pure YM to TMYM is not smooth as one would hope, because of different phase space geometries. From \eqref{eq:psipsi}, we have learned that transition from pure CS to TMYM requires splitting of CS into two pieces with half the level. This could have interesting applications in condensed matter physics.

CS splitting does not seem to appear if one uses the real polarization where $\psi=\psi[A_z, A_{\bar{z}}]$. This can be seen clearly in refs. \citen{Asorey1993477,Grignani1997360,Karabali:1999ef}. These earlier works study TMYM theory in a perspective where YM theory is perturbed by a CS term, while our work is exactly the opposite. Ideally, all polarizations should lead to the same inner product; but in this case, different polarizations create different mathematical difficulties that force one to focus on certain scales. For TMYM theory, holomorphic polarization facilitates studies near the CS limit, while the real polarization is suitable for the near YM limit\cite{nair}. The difficulties that arise when using the real polarization can be summarized as follows. First, $A_z$ and $ A_{\bar{z}}$ do not commute in the pure CS limit; the wave-functional cannot depend on both, at very large distances. This makes it impossible to check the CS limit of the wave-functional. But in the holomorphic polarization TMYM wave-functional reduces to CS wave-functional nicely. Second, in the real polarization, the wave functional cannot depend on $E$-fields. But in the near CS limit, $E$-fields dominate and $B$-fields are negligible. This creates a difficulty in studying large distances. Since CS splitting occurs at a scale where the first order $E$-field contributions are important, it is not surprising to not see this feature in a polarization that cannot resolve this scale. Similarly, holomorphic polarization is not helpful in studying smaller scales where $B$-field contributions are important, since $\psi[A_{\bar{z}},\tilde A_{\bar{z}}]$ does not depend on $B$-fields. Thus, to study the near CS limit, holomorphic polarization should be chosen. Therefore, our results cannot be compared\cite{nair} with the results of refs. \citen{Asorey1993477,Grignani1997360,Karabali:1999ef}, since they focus on a different scale where CS splitting does not occur.

At the end of our calculations, by writing \eqref{eq:equivalencegnrl} we showed that loop operator expectation values of CS and TMYM theories are related at large distances. The equivalence tells us that expectation values of both Wilson loops and 't Hooft loops in TMYM theory are equal to CS Wilson loop expectation values, up to a change in the level number. A more interesting result is that the expectation value of the product of these loops in TMYM theory is equal to the product of Wilson loop expectation values in CS theory. These results show that not only in the pure CS limit but also in the near CS limit, the observables of TMYM theory are link invariants. Both Wilson loops and 't Hooft loops can separately form links that satisfy the skein relation \eqref{eq:skeinw}, but a mixed link of these loops does not, even though it is still a link invariant.

One final important point is that taking $k$ to be a large integer does not seem to cause any problems in obtaining a topological theory at large distances. This indicates that having a large level number does not alter the existence of the mass gap in TMYM theory.

%-------------------------------------------------------------------------------------------------------------------------------------------
%-------------------------------------------------------------------------------------------------------------------------------------------

%\begin{center}
%\noindent\line(1,0){200}
%\end{center}
\newpage
\section*{Acknowledgements}
I am very grateful to Vincent Rodgers and Parameswaran Nair for suggesting this problem and for their supervision and support. I would like to thank Charles Frohman for his time and for helping me on knot theory. I thank Brian Hall for his help on geometric quantization. I thank Edward Witten, Tudor Dimofte and Chris Baesley for helpful comments. I also would like to thank the entire Diffeomorphisms and Geometry Research Group of the University of Iowa for useful comments and discussions. Finally, I would like to thank Suzanne Carter, Wade Bloomquist and Hart Goldman for their interest and early contributions. This work was partially supported by National Science Foundation grant NSF1067889.

%-------------------------------------------------------------------------------------------------------------------------------------------
%-------------------------------------------------------------------------------------------------------------------------------------------

%--------------------------------------------------------------------------------------

\begin{center}
\noindent\line(1,0){200}
\end{center}
\footnotesize
\bibliography{Bibliography.bib}
\bibliographystyle{unsrt}
%--------------------------------------------------------------------------------------

\end{document}